# Vulnerability Analysis of Digital Banks' Mobile Applications

*PV Falade, GB Ogundele

Department of Cyber Security, Nigerian Defence Academy, Kaduna

*Corresponding Email: pvfalade@nda.edu.ng

**ABSTRACT**

There is a rapid increase in the number of mobile banking applications users due to an increase in smart mobile devices. Mobile banking is a financial transaction and service offered through mobile devices. Almost all financial institutions now provide mobile banking services to their customers. However, the security of mobile banking applications is of huge concern because of the amount of personal data and information they collect. If an attacker gets hold of personal information, they can access bank payment or card accounts. This research aims to analyze the vulnerability of the UK digital banks' applications to identify vulnerabilities in the apps and proffer countermeasures that can help improve the security of the bank applications. Androbugs, a vulnerability scanner, was used to analyze the vulnerability of six digital banks' android applications. Starling, Monese, Atom bank, Transferwise, Monzo, and Revolut were scanned. All the scanned digital banks' applications have vulnerabilities; however, some have more vulnerabilities than others. For example, Revolut's mobile application has the highest number of identified vulnerabilities, while the Starling mobile application has the least identified vulnerabilities. Therefore, there is a need for more security in the digital banks' applications as well as other mobile banking applications.

## INTRODUCTION

There is a rapid increase in the number of mobile banking applications users due to an increase in smart mobile devices (Bucko, 2017; Luvanda, 2014; Palma et al., 2020; Szczepanik & Jóźwiak, 2018). Mobile banking is a financial transaction and service offered through mobile devices (Hayikader et al., 2016). Almost all financial institutions now provide mobile banking services to their customers. Some business reasons for embracing mobile banking services are attracting new customers, keeping existing customers, projecting market leadership, and meeting cost and competitive pressures (Bakar et al., 2017; Digital.ai, 2021; Nwoduh, 2019). While to the customers, mobile banking is a convenient, easy and fast way to perform transactions remotely (Szczepanik & Jóźwiak, 2018).

Although mobile banking services offer lots of advantages, we cannot deny the security challenges they pose to organizations that own them and customers that use them (Bucko, 2017; Digital.ai, 2021; Szczepanik & Jóźwiak, 2018). Security of mobile banking applications is of huge concern because of the amount of personal data and information they collect. If an attacker gets hold of personal information, they can access bank payment or card accounts (Bucko, 2017; Digital.ai, 2021). Therefore, most customers are still reluctant to use mobile banking apps because of the fear of fraudsters (Digital.ai, 2021).

Moreover, the insufficient security of mobile banking applications gives customers a justifiable reason to fear. Vulnerability in mobile applications is exposed by reverse engineering. 70% of digital banking fraudulent activities stem from the mobile channel by 2021's second-quarter—mobile banking app fraud amounting to the highest increase (Digital.ai, 2021). Cyber attackers usually develop scoundrel mobile applications from decompiled legitimate banking apps' source code (Digital.ai, 2021; El Janati El Idrissi et al., 2017) and upload to app stores; the rate of uploading reprobate mobile apps has reached 66% (Digital.ai, 2021). With the rate at which Mobile security risks are increasing everywhere, it is important to conduct a security assessment of mobile banking applications.

This research aims to assess the UK digital banks mobile applications to find the security vulnerabilities in the android apps, identify cyber threats that can exploit the vulnerabilities found on the digital banks' apps, and recommend countermeasures for some of the common vulnerabilities. The research shall use Android application vulnerability scanners to assess mobile apps. The scope of this research shall focus on the security of mobile banking applications, specifically android applications.

The remaining part of this article is grouped into five sections. Section 2 gives the literature review of related works, section 3 gives the methodology, the results of the security assessment with discussions are presented in section 4. While section 5 and 6 gives the conclusions and recommendations respectively.

## LITERATURE REVIEW

Digitalization is increasing globally, causing a great shift from traditional banks to digital banks, from physical to more digital. Fintech is a term that combines financial and technology. Fintech makes it possible for digital banks to operate without physical contacts. Bank customers can perform all activities with their smartphones or the computer through the inventions of Fintech. Digital banks have not only been a trend during recent years but have become even more appreciated and popular because of the COVID-19 pandemic. Many countries had to be on lockdown or cause disruption to the citizens' daily way of life. Digital banks which offer all of their services online enable people to stay at home while still performing financial transactions. Accessing the banking services through online channels became necessary to keep the country's economy growing (Nilsson & Lehmann, 2020).

Consequently, digital banks also called 'Challengers banks' are stealing the markets of standard banks. These banks focus more on growth than on profit by offering more value to customers through lowering of fees, streamlining processes posing as international banks and transacting in foreign countries without high fees and exchange rates. For example, opening an account with a digital bank takes a few minutes compared to the tedious complicated process with a traditional bank (Nilsson & Lehmann, 2020). Therefore, the rate at which these banks are growing is alarming. For instance, Revolut, one of the digital banks, has more than 8 million customers. Revolut was founded in 2015 and is already expanding to US and Australia. Likewise, Monzo has over 4 million customers in the UK. UK digital accounts now hold above 14% market share of primary accounts being switched. Nubank, a digital bank in Brazil, is now the largest digital bank globally, with more than 15 million unique customers (Deloitte, 2020). 71% of millennials would rather visit a dentist than a bank, according to the survey carried out by Millennial Disruption Index, and the millennial forms the target segment for banks. And this explains the speedy growth of digital banks without physical branches (Nilsson & Lehmann, 2020). The UK has been recognized as the global banking leader (PwC, 2018). For this reason, this research shall be focusing on the UK digital banks. UK digital banks are Starling, Monese, PAYSEND, cashplus, bunq, ANNA, Monzo, TANDEM, Atom, Transferwise, and Revolut.

Some people are comfortable with using digital banks. Still, others have no option but to use them because these banks tend to have lower requirements for opening an account compared to standard UK banks with high street physical branches. For example, an international student or a visitor in the UK without proof of address is forced to opt for digital banks for quick, convenience, and ease of use. Thereby, even though there is a low level of trust for such digital banks, people still use them. The relevance of digitalized banking will continue to the future. When transactions are made, no physical component is required. Traditional banks might be seen as an old way of banking, making them obsolete in the near future, with digital banks dominating. The physical interactions during digitalized banking might only be required during advice sections. Also, the advice section can be done through video calls as society is now familiar with the new way of communicating. Most physical meetings are now arranged through Zoom, Skype, Google meet, and others (Nilsson & Lehmann, 2020).

For this research, UK digital banks is defined as banks that live on smartphones (app-based) with no high street physical branches, but some have headquarters addresses. Some mobile accounts provide ATM (Automated Teller Machine) withdrawals or cash deposits, while others do not. Everything about banking with digital banks is done on mobile applications, from opening the account to managing the account (Nilsson & Lehmann, 2020). The security of these mobile applications is almost equivalent to the security of the physical high street branches of other standard banks. Mobile applications are not optional but compulsory for all users since there is no option to walk into the physical branch. These mobile applications need to be secure to ensure the safety of their customers.

Mobile applications can be defined as applications that run on mobile devices displaying information on the device's screen. Mobile banking applications can be either Android or iOS applications depending on the operating system used on a mobile device (ARGAW, 2018; et al., 2020). However, IBM Security Trustee Research discovered that more mobile fraud toolkits are offered in the underground forums. Some of the general risks to mobile banking applications and the use of mobile devices in general are: malicious code and applications, violation of privacy, payment infrastructure, wireless carrier infrastructure, and SMS vulnerabilities. Other threats include: eavesdropping, malware, lack of user awareness, third party application threat, phishing, platform malfunction, denial of service, unencrypted Wi-Fi, application/ device malfunction, unauthorized access, and loss theft or improper disposal of device. Top of the list of security issues of mobile banking applications is fraud and identity theft.

Szczepanik et al. (Szczepanik & Jóźwiak, 2018) investigated the security issues in mobile banking implementations through GSM (Global System for Mobile communication) network with South African banks as a case study. Their goal was to provide protocols for users to secure mobile banking applications via GSM network and GPRS (General Packet Radio Services) mediums (Szczepanik & Jóźwiak, 2018). Jozef (Bucko, 2017) analyzed the security of mobile devices in Slovakia's chosen banks, checking if the settings of the mobile devices are appropriate for using mobile banking applications securely as part of the work done (Bucko, 2017). Also, El Janati et al. (El Janati El Idrissi et al., 2017) presented some security concerns peculiar to android banking applications using reverse engineering (El Janati El Idrissi et al., 2017). On the other hand, Anthony (Luvanda, 2014) proposed a framework for protecting mobile banking apps from Man in the middle attack, ensuring that communications between application and server are secure.

Although there are existing tools for scanning the vulnerabilities of Android and iOS applications in general which are QARK (Quick Android Review Kit), Androbugs, and MobSF (Mobile Security Framework), Dexcalibur, StaCoAn, Runtime Mobile Security, Ostorlab, Quixxi, SandDroid, ImmuniWeb and App-Ray (Kumar, 2020), AUSERA (Automated Security Risk Assessment) was a framework specially developed by Chen et al. (Chen et al., 2020) to assess mobile banking applications. They used this framework to assess about 693 banking apps across the globe. They found AUSERA to outperform QARK, MobSF, and Androbugs in terms of precision rate and time. QARK has the best precision rate amongst the other three. During their empirical assessment, about 2,157 weaknesses were found across the 693 real-world banking apps (Chen et al., 2020). Although AUSERA is proven to be better in assessing mobile banking app than other scanners, it is not accessible for use. Therefore, this research used a general vulnerability scanner.

## METHODS

The security of the UK digital banks' mobile applications shall be assessed by scanning for vulnerabilities in the Android versions of mobile banking applications. The vulnerability scanner will identify security loopholes in each of the applications. Randomly selected six (6) UK digital banks; Starling, Monese, Monzo, Atom, Transferwise, and Revolut banks will be scanned using Androbugs. AUSERA specifically created for mobile banking applications could have been best fit but inaccessible. Also, attempt to use QARK on Windows came with many errors that we could not fix due to time constraints.

**Androbugs**

Androbugs is an open source vulnerability scanner developed by Yu-Cheng Lin and used by both hackers and developers to find potential security vulnerabilities in Android applications (Lin, 2015; Markiewicz, 2018). The framework has been published under the GNU General Public License v3.0 and written in python. It works on APK files with no need for root permissions on the device. However, it does not have an impressive Graphical User Interface (GUI) interface. The requirement for running Androbugs is to install Python 2.7.x on the intended device, while the PyMongo library is needed for massive analysis. The massive analysis involves running many applications using the Mongo database for analysis optimization and saving of results (Markiewicz, 2018). The massive analysis mode was not used for this dissertation since we scanned six (6) mobile applications. Androbugs Framework has the following features (Lin, 2015; Markiewicz, 2018):

- Search for security vulnerability
- Verify if the codes follow best practices
- Verify all dangerous shell commands
- Analyze more than one application at a time
- Verify the application's security protection

After the analysis by Androbugs, a report is automatically generated, and the report contains the following (Lin, 2015):

i. Title of vector
ii. Paths for source code
iii. Severity level of vulnerability (critical, warning, notice, info.)
iv. Category of vector
v. Vulnerability background knowledge
vi. Recommendations on how to mitigate the vulnerability

**AndroBugs Scanning Process**

The following steps were followed to scan the downloaded APK files of each digital bank mobile application using Androbugs on Windows:

i. Open the command prompt on windows and type "mkdir C:\AndroBugs_Framework" to create a directory.
ii. Change directory to AndroBugs_Framework using this command: "cd C:\AndroBugs_Framework."
iii. Unzip the latest Windows version of AndroBugs Framework from Windows releases.
iv. Go to Computer->System Properties->Advanced->Environment Variables. Add "C:\AndroBugs_Framework" to the "Path" variable.
v. Download the APK file of the application to be scanned.
vi. Open command prompt and type "androbugs.exe –h" click on "Enter' to run.
vii. Execute this command: "androbugs.exe -f [APK file]" to start analyzing the APK file.
viii. After the analysis, the output is downloaded.

**RESULTS**

**Mobile Banking Applications Vulnerabilities**

Security vulnerability of mobile banking applications discussed in this chapter are gotten from the identified critical issues, warnings and notices reported from the six scanned applications. Almost all the critical issues identified across the six applications are included as mobile banking application security vulnerabilities since they are confirmed vulnerability. In addition, some of the warnings reported across the six scanned applications are also included, but only a few of the notices were considered security vulnerabilities for mobile banking applications.

Vulnerability is a weakness in the design, implementation, operation, or internal control of a process that, when exposed, could be exploited by a threat from a threat event (Garrett, 2011). The vulnerabilities for mobile bank applications identified were picked based on their ability to be exploited by a threat event leading to harm. For example, security vulnerability such as "file unsafe deleting" is noticed but considered a

47

security vulnerability for mobile banking applications because of the sensitivity of data used in mobile banking applications. It will be risky for that sensitive information to be recovered in the case of lost or stolen devices. Also, screenshot capturing might not necessarily be a vulnerability to other applications that deal with less sensitive information. Still, screenshot capturing can lead to sensitive information leakage for mobile banking applications. A third party can easily copy vital information out of the application within the shortest time accessing the application (Chen et al., 2020). The identified security vulnerabilities for mobile bank applications are explained as follows. Note more vulnerabilities can be identified and added to the list, perhaps from scanning other digital bank applications outside these six.

1. **Implicit intent for service:** a messaging object can request action from another application component. On the other hand, a component that carries out operations without a user interface (in the background) is called a service. An implicit intent is a type of intent that does not specify the component but rather declares operations, generally allowing other applications' components to handle it (*Intents and Intent Filters*, 2022). It is a security hazard to use implicit intent to start a service in mobile banking applications. The service that will respond to the intent might not be ascertained, leaving the user ignorant of what service starts. Not all users are aware of this or know how to fix it, making it a vulnerability that an attacker can exploit.
2. **Misconfiguration of intent-filters:** another vulnerability related to intent is the misconfiguration in the "intent-filter" of these components. Config "intent-filter" should have at least one "action." If the intent filter is not configured properly, it can be exploited by a threat event causing harm to the application.
3. **Content Provider access from other apps on the device:** the repository data of an application is managed by a content provider. Although content providers are meant to be managed by other applications (*Content Provider Basics*, 2021), permissions should control the access. For example, it is a risk to allow other applications on the device to access the content provider of the mobile banking application without permission because there might be malicious applications installed on the user's device, which can exploit the mobile banking application if granted access.
4. **Remote code execution:** WebView "addJavascriptInterface" vulnerability can be used in allowing the host application to be controlled through JavaScript. This feature can be both powerful and risky because this method in a WebView with untrusted content could allow the malicious manipulation of host applications with permission through java code. Additionally, attackers can utilize this method to execute malicious code remotely.
5. **Getting IMEI and device ID:** getting the IMEI and device ID can be an issue with devices that do not have them. Also, giving out information about the users' devices might be prone to information leakage.
6. **Normal protection-level of permission:** The protection level of "normal" or default applications is not secured because it allows other applications to register and receive messages for the mobile banking application. It is a vulnerability because the message transmitted from or to the mobile banking application can be sensitive, and exposing it to other applications will be risky.
7. **Local file system access:** An attacker can inject malicious script, allowing the attacker to exploit other local resources in the device. For example, it means the attacker will use the mobile banking application to exploit the device's local resources.
8. **Webview Javascript enabled:** enabling javascript in Webview exposes it to cross-site scripting (XSS) attacks.
9. **Not executing 'root' or system privilege checks:** it is important for mobile banking applications to check the devices for "root" permission, mounting filesystem operations, or monitoring system before installation. Since some devices undergo "rooting" or "jailbreaking" by their users to gain higher privileges exposing the device to attacks. Therefore, not checking for "root" or system privilege is vulnerable.
10. **ADB backup:** ADB Backup is a tool for backing up all files. The risk associated with ADB backup in the case of mobile banking applications is the sensitivity of data used on the application. Also, when the backup falls into the hands of a third party or an attacker, it can lead to information leakage. Sensitive data found in ADB backup includes the lifetime access token, username or password, and other sensitive data about the mobile banking application.
11. **File unsafe deleting:** mobile application applications that allow unsafe deleting are vulnerable. Unsafe deleting means that everything deleted can be recovered by the user or an attacker, especially for rooted devices. For example, if the phone is sold or stolen, the new user will recover sensitive information.
12. **Not checking package signature code:** checking the package signature in the code helps check if a hacker hacks the application. Not making provisions to check for package signature code is vulnerability because the application might be hacked and unnoticed.
13. **Allowing Screenshot capturing:** screenshot is an easy way of harvesting users' sensitive information from the application. Although not a vulnerability to all applications but a vulnerability to mobile banking applications because of the nature of data it handles. The data used in mobile banking applications are too sensitive. Allowing screenshot capturing is a vulnerability because a third party or criminal can easily copy sensitive information out of the application within the shortest possible time of access.
14. **Not checking APK installer sources:** criminals are making fake, fraudulent applications and making them available on the internet. These fake applications look like the original but have a malicious intention. Therefore, it is important to check for the source of the APK installer, and failure to do so is a vulnerability to mobile banking applications.

There are fourteen (14) identified security vulnerabilities for mobile banking applications. Table 1 shows the security vulnerabilities in each of the six scanned applications; colour green shows that there is no vulnerability while the colour yellow shows there is vulnerability.

**Table 1: Security Vulnerabilities**

| Security Vulnerability | Starling | Monese | Atom | Transferwise | Monzo | Revolut |
|---|---|---|---|---|---|---|
| Implicit intent for service | green | yellow | green | green | green | yellow |
| Misconfiguration of intent-filters | green | green | green | green | green | yellow |
| Content Provider access from other apps on the device | green | green | green | green | green | yellow |
| Remote code execution | green | yellow | green | green | green | green |
| Getting IMEI and Device ID | green | yellow | green | yellow | green | green |
| Normal protection-level of permission | green | green | green | green | green | yellow |
| Local file system access | green | yellow | yellow | yellow | yellow | yellow |
| Webview JavaScript enabled | green | yellow | yellow | yellow | green | yellow |
| Not executing 'root' or system privilege checks | green | green | yellow | yellow | yellow | yellow |
| ADB backup | yellow | green | green | green | green | green |
| File unsafe deleting | yellow | yellow | yellow | yellow | yellow | yellow |
| Not checking Package signature code | green | green | green | green | green | yellow |
| Allowing screenshot capturing | yellow | yellow | green | yellow | green | yellow |
| Not checking APK installer sources | 🟢 | 🟢 | 🟡 | 🟢 | 🟡 | 🟡 |

**Comparison of the Scanned Mobile Applications Vulnerabilities**

Out of the total security vulnerabilities for mobile banking applications identified from scanning the six digital banks' mobile applications, Starling bank contains the least number of vulnerabilities. On the other hand, the Revolut bank mobile application has the highest vulnerabilities. Atom, Transferwise, and Monzo contain the same number of vulnerabilities, although not the same vulnerabilities. The percentages of the security vulnerabilities contained in the different applications are shown in Table 2.

From Table 1, all the scanned applications have the vulnerability "file unsafe deleting." Also, all applications except the Starling bank mobile application have the vulnerability "local file system access." Only Revolut bank mobile application has the following vulnerabilities; "misconfiguration of intent-filters," "content provider access from other apps on the device," "normal protection-level of permission," and "not checking package signature code." Also, only the Starling bank mobile application has the vulnerability "ADB backup." Finally, the vulnerability "remote code execution is found in only Monese bank mobile application."

Furthermore, from Table 1, it is seen that the vulnerability "implicit intent for service" is contained in only Monese and Revolut bank mobile applications; the other four applications use explicit intent for service. Also, Monese and Transferwise bank mobile applications get IMEI and device ID; others do not. Webview Javascript is disabled for other mobile applications except for Monese, Atom, and Transferwise bank mobile applications, which have their Webview JavaScript enabled. Atom, Transferwise, Monzo, and Revolut bank mobile applications do not check for "root" or system privilege execution, while Starling and Monese bank mobile applications perform checking. Screenshot capturing is prevented in Atom and Transferwise bank mobile applications while Starling, Monese, Monzo, and Revolut permit screenshot capturing. Finally, Monzo, Atom, and Revolut bank mobile applications do not have a code for checking APK installer sources while Starling, Monese, and Transferwise perform checks.

**Table 2: Security Vulnerability Percentages**

| Digital Banks | Total number of security vulnerabilities identified in the application | Percentage of security vulnerabilities identified |
|---|---|---|
| Starling | 3 | 21.43% |
| Monese | 7 | 50.00% |
| Atom | 5 | 35.71% |
| Transferwise | 5 | 35.71% |
| Monzo | 5 | 35.71% |
| Revolut | 10 | 71.43% |

**Threats and Countermeasures**

In this section, threats that can exploit the identified vulnerabilities are presented. And to prevent an attack from occurring, the countermeasures that can prevent the threat from exploring the vulnerabilities are also provided.

**Threats**

Threats are potential harm that can exploit vulnerabilities to cause an attack (Garrett, 2011). The threats that can exploit the mobile applications, the vulnerability they exploit, and a description is presented in Table 3:

- Lack of user awareness
- Application malfunction
- Third-party application threat
- Unauthorized access
- Information Leakage
- Malware
- Cross-site scripting threat
- Platform manipulation
- Improper disposal of the device
- Hacking
- Phishing through fake applications

**Table 3: Description Of Threats**

| Security vulnerability | Threat | Description |
|---|---|---|
| Implicit intent for service | Lack of user awareness | Most users are not aware of the different intents for services, so if the application is not correctly set, the user will not know it. |
| Misconfiguration of intent-filters | Application malfunction | Without properly configuring the application, unexpected actions can occur. |
| Content Provider access from other apps on the device | Third-party application threat | There might be malicious applications on the device, and gaining access without permission could be harmful to the mobile banking application. |
| Remote code execution | Unauthorized access | If remote code execution is allowed, an attacker can access the application remotely without authorization. |
| Getting IMEI and Device ID | Information leakage | Getting IMEI and device ID might lead to information leakage. |
| Normal protection-level of permission | Third-party application threat | If the protection-level permission is not set, others can register and receive messages for the application. |
| Local file system access | Malware (malicious code) | Malicious codes can be injected, thereby exploiting the system. |
| Webview JavaScript enabled | Cross-site Scripting threat | It makes it prone to cross-site scripting attacks. |
| Not executing 'root' or system privilege checks | Platform manipulation | Sometimes users "root" or "jailbreak" devices to gain higher privileges. |

| | | Unfortunately, this can leak sensitive information from the application, so it is important to check for 'root' in systems where the mobile banking apps are installed. |
|---|---|---|
| ADB backup | Improper disposal of the device | Improper device disposal with the installed application can lead to ADB backup falling into the wrong hands. |
| File unsafe deleting | Improper disposal of devices | If the device is lost, sold, or stolen, the deleted sensitive information can be retrieved. |
| Not checking Package signature code | Hacking | When hacking occurs, it cannot be detected by the application. |
| Allowing Screenshot capturing | Improper disposal of the device | If the mobile banking application gets into the wrong hand, the attacker can easily screenshot sensitive information out of the application. |
| No APK installer sources checks | Phishing through fake applications. | Criminals are fond of creating fake applications to deceive users into installing them, thereby stealing sensitive information from the user. Therefore, it is important to check the APK installer source to ascertain it is genuine. |

## Countermeasures

Countermeasures are security mechanisms implemented to prevent attacks by either eliminating the vulnerability or preventing the threat occurrence. In this section, two groups of countermeasures shall be presented, the countermeasures to be implemented by the developers to fix the vulnerability and those to be implemented by the users to prevent the likelihood of threat occurrence.

### Developer's Countermeasures

The developers need to make changes in the code of applications to fix the vulnerabilities. Table 4 gives the vulnerabilities and their countermeasures.

**Table 4: Countermeasures for Developers**

| Security vulnerability | Countermeasures |
|---|---|
| Implicit intent for service | Always use explicit intent when starting a service. |
| Misconfiguration of intent- | Config "intent-filter" should have at least one |
| filters | "action." |
| Content Provider access from other apps on the device | Set at least "signature" protectional Level permission or make it "false." |
| Remote code execution | Modify code to disallow remote code execution. |
| Getting IMEI and Device ID | If the device ID is needed, use the "Installation" framework instead. |
| Normal protection-level of permission | The app should declare the permission with the "android:protectionLevel" of "signature" or "signatureOrSystem" so that other apps cannot register and receive messages for this app. android:protectionLevel="signature" ensures that apps with request permission must be signed with the same certificate as the application that declared the permission. |
| Local file system access | This can be mitigated or prevented by disabling local file system access. (It is enabled by default). |
| Webview JavaScript enabled | Disable Webview Javascript. |
| Not executing 'root' or system privilege checks | There should be a code for checking for "root" or system privilege in the device. |
| ADB backup | Disable ADB backup in an application. |
| File unsafe deleting | Do not use "file.delete()" to delete essential files. |
| Not checking Package signature code | There should be a code in the application for checking package signature. |
| Allowing Screenshot capturing | This application should have a code for preventing screenshot capturing. |
| No APK installer sources checks | The application should have a code for checking APK installer sources. |

**User's Countermeasures**

The countermeasures in this section are the responsibilities of the users to implement to reduce the likelihood of occurrence of threats.

- **Do not use mobile banking applications for jailbreak or rooted mobile devices:** this will help prevent the threat of platform manipulation.
- **Download mobile banking applications from trusted app stores:** doing this will help reduce the risk of phishing through fake applications.
- **Use mobile anti-virus applications:** using updated anti-virus or anti-malware on the mobile phone can help prevent the execution of malicious codes on the device.
- **Physically protect the mobile device:** physically protecting the mobile device keeps it safe from criminals and third parties, thereby reducing the threat of improper disposal of devices that can lead to information leakage.
- **Update mobile banking applications regularly:** this will help the user have a better secured and updated version. Moreover, the updated version usually has some of the vulnerabilities fixed. Therefore, the likelihood of occurrence of threats like information leakage, cross-site scripting threat, unauthorized access, third party applications threat, and application manipulation will be reduced.

- ❖ **Update mobile device's operating system:** this will help tackle threats like hacking and application malfunction.

## CONCLUSIONS AND RECOMMENDATIONS

In conclusion, there is a great need for more security in the digital banks' Android applications. This research is beneficial to the digital banks, their developers, and users. All the scanned digital banks' applications have vulnerabilities; however, some have more vulnerabilities than others. For example, Revolut's mobile application has the highest number of identified vulnerabilities, while the Starling mobile application has the least identified vulnerabilities. Therefore, there is a need for more security in the digital banks' applications as well as other mobile banking applications.

For future work, other vulnerability scanners can scan more mobile banking applications to identify more vulnerabilities, thereby identifying more threats and recommending more countermeasures to users and developers of mobile banking applications. As a result, creating a more secure digital banking world for users.